\documentclass[12pt,preprint]{aastex}

\usepackage{epstopdf, graphicx, natbib}

\bibpunct{(}{)}{,}{a}{}{;}

\def\lya{\ifmmode {\rm Ly}\alpha~ \else Ly$\alpha$~\fi}
\def\lyan{\ifmmode {\rm Ly}\alpha \else Ly$\alpha$\fi}
\def\lyb{\ifmmode {\rm Ly}\beta~ \else Ly$\beta$~\fi}
\def\lyg{\ifmmode {\rm Ly}\gamma~ \else Ly$\gamma$~\fi}

\def\civ{\ifmmode {\rm C}\,{\sc iv}~ \else C\,{\sc iv}~\fi}
\def\civn{\ifmmode {\rm C}\,{\sc iv}~ \else C\,{\sc iv}\fi}

\def\cvin{\ifmmode {\rm C}\,{\sc vi} \else C\,{\sc vi}\fi}

\def\ovin{{{\rm O}\,{\sc vi}}}
\def\ovii{{{\rm O}\,{\sc vii}~}}
\def\oviii{{{\rm O}\,{\sc viii}~}}
\def\ovin{{{\rm O}\,{\sc vi}}}

\def\neix{{{\rm Ne}\,{\sc ix}~}}

\def\fevii{{{\rm Fe}\,{\sc vii}~}}

\def\fevii-xii{{{\rm Fe}\,{\sc vii-xii}~}}

\def\fexvii{{{\rm Fe}\,{\sc xvii}~}}

\def\sixiii{{{\rm Si}\,{\sc xiii}~}}
\def\sixiv{{{\rm Si}\,{\sc xiv}~}}
\def\sixivn{Si\,{\sc xiv}~}

\def\mgxii{{{\rm Mg}\,{\sc xii}~}}

\def\chandra{{\it Chandra}~}
\def\xmm{{\it XMM-Newton}~}
\def\suzaku{{\it Suzaku}~}

\def\ark564{{\it Ark~564}}
\def\mrk590{{\it Mrk~590}}
\def\ngc3783{{\it NGC~3783}}

\title{Detection of high velocity outflows in Seyfert 1 \mrk590}

\author{ A.~Gupta\altaffilmark{1,2}, S.~Mathur\altaffilmark{2}, 
Y.~Krongold\altaffilmark{3}}

\altaffiltext{1}{Department of Biological and Physical Sciences, 
                      Columbus State Community College, 
                      Columbus, OH 43215, USA;
                      agupta1@cscc.edu}

\altaffiltext{2}{Department of Astronomy, 
		The Ohio State University, 
		140 West 18th Avenue, 
		Columbus, OH 43210, USA}

\altaffiltext{3}{Instituto de Astronomia,
                 Universidad Nacional Autonoma de Mexico           
                 Mexico City, (Mexico)}

\begin{document}

\begin{abstract}

We report on the detection of ultra-fast outflows in the Seyfert~1
galaxy \mrk590. These outflows are identified through highly
blue-shifted absorption lines of \oviii \& \neix in the medium energy
grating spectrum and \sixiv \& \mgxii in the high energy grating
spectrum on board \chandra X-ray observatory. Our best fit
photoionization model requires two absorber components at outflow
velocities of 0.176c and 0.0738c and a third tentative component at
0.0867c. The components at 0.0738c and 0.0867c have high ionization
parameter and high column density, similar to other ultra-fast outflows
detected at low resolution by Tombesi et al. These outflows carry
sufficient mass and energy to provide effective feedback proposed by
theoretical models. The component at 0.176c, on the other hand, has low
ionization parameter and low column density, similar to those detected
by Gupta et al. in Ark~564. These absorbers occupy a different locus on
the velocity vs. ionization parameter plane and have opened up a new
parameter space of AGN outflows. The presence of ultra-fast outflows in
moderate luminosity AGNs poses a challenge to models of AGN outflows.

\end{abstract}

\section{Introduction}

Outflows are ubiquitous in AGNs, manifested by high-ionization
absorption lines in X-rays and UV \citep[and references
therein]{Mathur1995, Mathur1997}.  The X-ray absorbers, commonly known
as warm absorbers (WAs), have typical outflow velocities of
$100-1000~$km~s$^{-1}$. The discovery of ultra-fast outflows (UFOs) at
velocities $v \sim 0.1c$ in the hard X-ray band has added an intriguing
aspect to the rich field of AGN outflows.  These outflows are exhibited
by blueshifted absorption lines, produced mostly by highly ionized iron
\citep[and references therein]{Pounds2003a, Pounds2003b, Tombesi2010}.
The mass outflow rate of UFOs can be comparable to the accretion rate
and their kinetic energy can correspond to a significant fraction of the
bolometric luminosity \citep{Tombesi2013, Pounds2003a, Pounds2003b,
Reeves2009}.  Thus these outflows can provide effective feedback that is
required by theoretical models of galaxy formation
\citep[e.g.,][]{Hopkins2010, Silk1998}.

While WAs are detected unambiguously in the high-resolution soft X-ray
spectra and their physical properties and kinematics are well determined
\citep[e.g.,][]{Krongold2003}, the same cannot be said about UFOs.  The
UFOs in $\textit{PG 1211+143}$ and $\textit{PG 0844+349}$
\citep{Pounds2003a, Pounds2003b} were detected in high-resolution
grating spectra, but the rest of the UFOs are detected in low-resolution
CCD spectra, observed with the \xmm and \suzaku. The rest frame energy
of the Fe-K$\alpha$ line transition is 6.38~keV, and at outflow
velocity of $\approx$ 0.1c the UFOs are observed at about 7.0 keV. This
falls in the region of the spectrum where the effective area and
resolution are low. As a result the significance of the absorption line
detection in the hard X-ray band is often questioned
\citep[e.g.,][]{Laha2014} and with only a few lines observed, accurate
parameterization of the photoionized plasma becomes difficult. These
difficulties were alleviated with our discovery of relativistic outflows
in the soft X-ray band in high-resolution spectra.

We recently discovered relativistic outflows in \chandra HETG spectra of
Seyfert~1 galaxy \ark564 \citep{Gupta2013b}; these detections are
robust, with high statistical significance. We identified highly
blueshifted absorption lines of \ovin, \ovii and \oviii and successfully
fitted them with a two-component photo-ionization model. The detection
of multiple lines at the same velocity makes the identification of the
relativistic outflow robust. This opens up an exciting new opportunity
of probing the relativistic disk winds in the soft X-ray band where we
have the best diagnostic power from multiple lines of multiples
ionization states of several elements and where the gratings response is
the best.

Excited by the discovery of relativistic outflows in \ark564, 
we looked for their presence in other AGNs; could it be that they 
were not found because nobody looked for them? The Chandra HETGS data of 
\mrk590 are tantalizing. \mrk590 is a bright Seyfert 1 galaxy 
(recently appears to have changed from Type 1 to Type $\sim1.9-2$; 
Denney et al. 2014) at $z=0.026$ that has been observed by the 
Einstein observatory \citep{Kriss1980}, the HEAO~1 \citep{Piccinotti1982}, 
the {\it BATSE} \citep{Malizia1999} and recently by \chandra and \xmm. 
\citet{Gallo2006} first reported the presence of a strong Fe~K emission line 
revealed in the 2002 $\sim10~$ks \xmm EPIC data. Later 
\citet{Longinotti2007} presented its complex spectrum with a reflection 
component using the $\sim100~$ks observations with \chandra and \xmm and 
confirmed the presence of the Fe K$\alpha$ emission line. Here we present a 
detailed reanalysis of the \chandra HETG observation of \mrk590. 

\section{Observation and data reduction}

\mrk590 was observed by \chandra High Energy Transmission Grating (HETG)
in 2004 for $\sim100~ks$ (ObsID: 4924). The HETG is comprised of two
gratings: the medium energy gratings (MEG) and the high energy gratings
(HEG), which disperse spectra into positive and negative spectral
orders. We reduced the data for both gratings using the standard Chandra
Interactive Analysis of Observations (CIAO) software (v4.3) and Chandra
Calibration Database (CALDB, v4.4.2) and followed the standard Chandra
data reduction threads\footnote{\url{http://cxc.harvard.edu/ciao/
threads/index.html}}. To increase S/N we co-added the negative and
positive first-order spectra and built the effective area files (ARFs)
using the \emph{fullgarf} CIAO script.

\section{Spectral Analysis}

For the spectral analysis, we binned the MEG and HEG spectra to their 
resolution element of $0.025~\AA$ and $0.012~\AA$ respectively and analyzed 
using the CIAO fitting package \emph{Sherpa}. Throughout this paper 
the quoted errors correspond to $1~\sigma$ significance. 

\subsection{Continuum Modeling}

We fit the intrinsic continuum of the source with an absorbed (Galactic
$N_{H}=2.7\times10^{20}~cm^{-2}$; Dickey \& Lockman 1990) powerlaw; the
fit parameters are reported in the Table 1. In the energy band of
$0.3-10~keV$, the integrated flux is
$9.1\times10^{-12}~ergs~s^{-1}~cm^{-2}$, which corresponds to an
unabsorbed X-ray luminosity of $1.4\times10^{43}~ergs~s^{-1}$.


Although absorbed powerlaw fits the continuum well 
(MEG: $\chi^{2}= 729, d.o.f.=679$, HEG: $\chi^{2}= 1120, d.o.f.=1199$), 
there are residual absorption line-like features in the spectral regions of 
$10-20~$\AA\ and $4-9~$\AA\ in the MEG and HEG
spectra respectively (Fig. 1 \& 2).  In the subsequent sections we
discuss the identification, statistical significance and possible
origin of these absorption lines.

\subsection{Identification of the Absorption Lines}

\subsubsection{Medium Energy Grating Spectrum}

In the MEG spectrum of \mrk590 the most prominent absorption feature ($>
3\sigma$; Fig. 3) is present at $16.03$~\AA. Addition of a gaussian
absorption feature (at $16.033\pm0.004$) to continuum model improves the
fit ($\Delta \chi^{2}= 17$ for 2 fewer d.o.f.) at the confidence level of
99.97\% according to the F-test \citep{Bevington1992}. The measured
equivalent width (EW) is $46.50\pm11.04~$m\AA.  Errors are at $1\sigma$
confidence level and are calculated using the ``projection'' command in
{\it Sherpa}.  According to the Gaussian probability distribution the
probability of detecting the line by chance is $1.3 \times
10^{-5}$. There are 681 wavelength bins in the MEG spectrum, so the
probability of finding an absorption line at the observed significance
anywhere in the spectrum due to random statistical fluctuations is
0.88\%. Thus it is highly unlikely that this absorption feature is due
to random statistical fluctuations.

We tried to identify the absorption feature detected at
$16.033\pm0.004~$\AA.  There is no known instrumental feature near this
energy (Chandra Proposers Observatory Guide, or POG). Though this line
could be associated with $z=0$ \oviii K$\beta$ line
($\lambda_{rest}=16.007~$\AA), we rule out this possibility because
there is no corresponding \oviii K$\alpha$ line
($\lambda_{rest}=18.967$\AA). Assuming this feature is produced by
\fexvii ($\lambda_{rest}=15.015$\AA) ions in an intervening warm-hot
intergalactic medium \citep[WHIM; ][]{Mathur2003, Nicastro2005}, the
system redshift would be $z_{WHIM}=0.066$ which is greater than the
$z_{AGN}=0.02638$.  Thus this feature could not be from an intervening
WHIM system and no other intervening absorption line is likely
either. Therefore we assume that this absorption line is intrinsic to
the source.

We identify this line based on a combination of chemical abundance, line
strength and assuming a very broad range of inflow/outflow velocities of
$\pm60,000~$ km~s$^{-1}$. The likely candidates for the $\sim16.033\AA$
feature are \fexvii ($\lambda_{rest}=15.015~\AA$;
$\lambda_{s}=15.411~\AA$, wavelength at the source redshift) with inflow
velocity of $\sim0.04c$, or \oviii ($\lambda_{rest}=18.96\AA$;
$\lambda_{s}=19.460~\AA$) with outflow velocity of $0.176\pm0.001c$. To
distinguish between the two possibilities of inflow and outflow, we
search for possible associations of other lines. We do not find any
possible association for inflows but found a 2.8$\sigma$ line at
$11.39\pm0.01$ corresponding to \neix ($\lambda_{rest}=13.447\AA$;
$\lambda_{s}=13.802~\AA$, Fig. 3) at the similar outflow velocity of
$0.175\pm0.001c$.

\subsubsection{High Energy Grating Spectrum}

In the \mrk590 HEG spectrum, we detected two absorption features at
$5.80~\AA$ and $5.88~\AA$ (Fig. 4) with EW of $25.0\pm4.4~$m\AA~ and
$21.4\pm3.2~$m\AA~ respectively.  The HEG and MEG overlap in this
wavelength range, so if real, these lines should also be present in the
MEG spectrum. We searched for these absorption features in the MEG
spectrum and found one at $5.91\pm0.04~$\AA with EW of
$25.3\pm12.5~$m\AA; this is consistent with the $5.880\pm0.001~$\AA HEG
line, though the error on the line EW in the MEG spectrum is
large. There is no absorption feature in the MEG spectrum corresponding
to the $5.80~\AA$ HEG feature, but again the EW limit in the MEG
spectrum is large, $13.13$m\AA, so the discrepancy is less than
$2\sigma$. In the following we will treat both of these features as real
lines, but keeping in mind that our confidence in the $5.80~\AA$ feature
is not as strong as that in the $5.88~\AA$ feature.

In the HEG spectrum, both lines are present at high significance with a
very low probability of finding them by chance ($3\times 10^{-4}$\% and
$6\times 10^{-5}$\%) anywhere in the spectrum.  The only possible
identification we found for them is of \sixiii and \sixiv at outflow
velocities of $0.085c$--$0.067c$. We will see later that the
photo-ionization model supports the identification of these lines as
\sixivn.

\section{Photo-ionization model: fitting with PHASE}

To test the validity of our tentative identifications of absorption
features with high-velocity outflows, we fitted them with our
photo-ionization model code PHASE \citep{Krongold2003}. The code has 4
free parameters per absorbing component, namely 1) the ionization
parameter $U=\frac{Q(H)}{4\pi r^{2}n_{H}c}$; where \emph{Q(H)} is the
rate of H ionizing photons, \emph{r} is the distance to the source,
\emph{$n_{H}$} is the hydrogen number density and \emph{c} is the speed
of light; 2) the hydrogen equivalent column density \emph{$N_{H}$}; 3)
the outflow velocity of the absorbing material \emph{$V_{out}$} and 4)
the micro-turbulent velocity \emph{$V_{turb}$} of the material.  PHASE
has the advantage of producing a self-consistent model for each
absorbing component, because the code starts without any prior
constraint on the column density or ionization fraction of any ion. We
have assumed solar elemental abundances \citep{Grevesse1993} in the
following analysis.

The residual features at $16.033\pm0.004$ and $11.39\pm0.01$ in the MEG
spectrum are well fitted with one PHASE component of $\log U=
-0.10\pm0.25$ and $\log N_{H}= 20.2 \pm 0.2$ and outflow velocity of
$0.176\pm0.001 c$ (Fig. 5) and the lines are indeed \oviii K$\alpha$ and
\neix. The addition of the PHASE component to the continuum model
significantly improves the fit ($\Delta \chi^{2}= 15$ for 3 fewer d.o.f.). 
According to F-test, the absorber is
present at a confidence level of 99.66\%. We call this component as the
“high-velocity--low-ionization-phase (HV-LIP)” absorber.  Our best fit
absorber model also predicts absorption from \ovii at $18.26~$\AA. As
shown in figure~5, the data are consistent with the prediction. However,
to securely detect the \ovii feature, we require higher signal to noise
data.

To model the HEG absorption features (at $5.08~\AA$ and $5.88~\AA$)
identified as \sixiii or \sixivn, required two PHASE components with
$\log U= 2.5\pm0.3$ and $\log N_{H}= 23.50\pm0.03 $ at different outflow
velocities (Fig. 6). The photo-ionization model fitting favors the
identification of these features as ionic transition of \sixiv produced
in highly ionized medium outflowing with velocity of $0.0867\pm0.0007c$
and $0.0738\pm0.0004c$ respectively. Both the absorbers are required at
high significance of 99.999\% ($\Delta \chi^{2}= 24$ for 4 fewer d.o.f.)
and 99.996\% ($\Delta \chi^{2}= 28$ for 4 fewer d.o.f.) respectively.
The HEG best fit model with two absorbers also requires \mgxii absorption
features at $7.90$ and $8.01~\AA$; these lines are in fact present and
are well fitted by the model. The best fit PHASE parameters are reported
in Table 2 and the best fit models are shown in Fig. 6. These absorbers
fit the high ionization lines, so we will refer to these absorbers as
the “high-velocity--high-ionization phase I (HV-HIP-I)” and
“high-velocity--high-ionization phase II (HV-HIP-II)” components.
Successfully modeling the residuals in the MEG and HEG spectra further
supports the presence of high-velocity outflows in \mrk590.

\section{Mass and Energy ouflow rate estimates}

Without knowing the location of the absorber, its mass and energy
outflow rates cannot be well constrained. The lower limit on the
absorber location can be determined assuming that the observed outflow
velocity is the escape velocity at the launch radius $r$:
i.e. $r=\frac{2GM_{BH}}{v_{out}^{2}}$.  \citet{Peterson2004} determine
the central black hole mass of \mrk590 to be $M_{BH}=(4.75\pm0.74)
\times 10^{7}M_{\odot}$. Using the above expression we get the minimum
distance of \mrk590 HV-LIP and HV-HIP absorbers of $\sim 1.5 \times
10^{-4}~$pc and $\sim 6.1 \times 10^{-4}~$pc respectively. This is 0.17
and 0.72 light-days from the nuclear black hole, and inside the broad
emission line region.

For a bi-conical wind, the mass outflow rate is $\dot{M}_{out} \approx
1.2 \pi m_{p} N_{H} v_{out}r$ \citep{Krongold2007}.  Substituting $r$
with $r_{min}$ and using outflow velocities of 52800~km~s$^{-1}$ and
26010~km~s$^{-1}$, we obtain lower limit on mass outflow rates of
$\dot{M}_{out} \geq 3.8 \times 10^{-5}~$M$_{\odot}~$s$^{-1}$ and
$\dot{M}_{out} \geq 1.5 \times 10^{-1}~$M$_{\odot}~$s$^{-1}$,
respectively.  These corresponds to kinetic luminosity of LIP and HIP
high velocity outflows of $\dot{E}_{K}\geq 1.1 \times
10^{41}~$erg~s$^{-1}$ and $\dot{E}_{K}\geq 2.2 \times
10^{44}~$erg~s$^{-1}$, respectively.  The kinetic luminosity of HIP high
velocity outflows is much larger than X-ray (2-10~keV) luminosity of
$7.0 \times 10^{42}~$erg~s$^{-1}$.  Thus in principle these outflows
have the potential to provide effective feedback.

\section{Discussion}

The  \chandra observation of \mrk590 has revealed low- and high-ionization 
high-velocity outflows. The low-ionization and low-column (HV-LIP) 
component is similar to those recently discovered in 
another Seyfert-type moderate luminosity AGN \ark564 (Gupta et al. 2013b). 
However, the outflows identified as HV-HIP-I and HV-HIP-II have parameters 
similar to that of UFOs (highly ionized $log\xi = 3 - 6~$erg~s$^{-1}$ and 
with large column densities $N_{H}=10^{22}-10^{24}~$cm$^{-2}$). 

Recently \citet{Tombesi2013} presented the connection between UFOs and
soft X-ray WAs. They strongly suggest that these absorbers represent
parts of a single large-scale stratified outflow and that they
continuously populate the whole parameter space (ionization, column,
velocity), the WAs and the UFOs lying at the two ends of the
distribution (Fig. 7). The \ark564 low-velocity WAs \citep{Gupta2013a}
and UFOs \citep{Papadakis2007} were in agreement with the linear
correlation fits from Tombesi et al., but the low-ionization low-column
high-velocity outflows in \ark564 probe a completely different parameter
space (\emph{U, $N_{H}$, $v_{out}$}), as shown in Fig.7. Similarly the
\mrk590 HV-HIP outflows probe the same parameter space as of UFOs, but
HV-LIP outflows probe the parameter space not covered by either
low-velocity WAs or high-ionization UFOs.

The presence of different absorption systems with different outflow
velocities may suggest that we are looking at the flows in a transverse
configuration with respect to the line of sight. This would avoid
possible collision between different components, that would result in
shock-heating of the gas and a decrease on the measured opacity. The
geometry would have to be conical and the flows must form a large angle
with the accretion disk. It is also expected that the orthogonal
component of the velocity might have a significant contribution to the
total velocity of the flow, but we only observe the component in the
line of sight. This would mean that the flows are even faster than
detected. In such a steady state configuration, the absorption lines
could be expected to remain constant over time, as we would always
observe the same section of the wind.

Alternatively, we may be seeing a radial flow, with a nearly spherical
geometry and a large velocity gradient in the line of sight. The
different velocity components could be explained if they are parts of
this single flow, formed by clumps in the flow. In this scenario, strong
variability in the observed absorption lines is expected, as the clumps
move further out from the central engine, and their velocity and
ionization state change. Observations of \mrk590 over time would allow
us to discriminate between these two scenarios.

As discussed in \citet{Gupta2013b}, the relativistic outflows cannot be
explained by simple models of radiation pressure driven winds.  Their
velocities are too large for their luminosities and
magneto-hydrodynamics may be involved. Magnetic fields are also
important for launching jets; could relativistic outflows be failed
jets? To get the in-depth view of the AGN outflows, its important to
search for more of such systems probing the complete parameter space.

\section{Summary}

We have detected high-velocity outflows in the Seyfert~1 galaxy \mrk590.
These absorbers are identified through highly blue-shifed absorption
lines of \oviii \& \neix in the MEG spectrum and \sixiv \& \mgxii in the
HEG spectrum. Our best fit photoionization model requires three absorber
components at outflow velocities of 0.176c, 0.0867c and 0.0738c. The
HV-LIP absorber has low -ionization and -column density while two HV-HIP
absorbers have high -ionization and -column density. All the absorbers
are required at high significance of 99.66\%, 99.999\% and 99.996\%
respectively. The HV-HIPs have sufficient mass and energy to provide
effective feedback proposed by theoretical models. However, the presence
of such UFOs in moderate-luminosity AGNs poses a challenge to models of
AGN winds.

\clearpage

\bibliography{apj}

\clearpage

\begin{deluxetable}{lccc}
\tabletypesize{\scriptsize}
\tablecaption{Continuum model parameters for the Mrk590 \chandra HETG spectra}
\tablewidth{0pt}
\tablehead{
\colhead{} & \colhead{Units} & \colhead{MEG} & \colhead{HEG}
}
\startdata
\textbf{Powerlaw}\\
Photoindex ($\Gamma$) &   &  $1.57\pm0.03$ &  $1.56\pm0.03$  \\
Normalization & $10^{-4}~ph~keV^{-1}~s^{-1}~cm^{-2}$ & $15.84\pm0.25$ & 
$13.36\pm0.37$ \\
\enddata
\end{deluxetable}

\clearpage

\begin{deluxetable}{lccccc}
\tabletypesize{\scriptsize}
\tablecaption{Absorption lines identified with high velocity outflows in 
the HETGS Spectrum of Mrk 590.}
\tablewidth{0pt}
\tablehead{
\colhead{Ion} & \colhead{$\lambda_{obs}$} & \colhead{$\lambda_{rest}$} & 
\colhead{$v_{out}$} & \colhead{$EW$}\\
\colhead{} & \colhead{$\AA$} & \colhead{\AA} & \colhead{km~s$^{-1}$} & 
\colhead{m$\AA$}}
\startdata
\textbf{Medium Energy Grating}\\
\oviii $K\alpha$  &  $16.033\pm0.004$  & $18.969$   &    $52936\pm62$   &  
$46.50\pm11.04$  \\
\neix             &  $11.39\pm0.01$  & $13.447$   &    $52423\pm217$   &  
$20.7\pm7.48$  \\
\textbf{High Energy Grating}\\
\sixiv &  $5.804\pm0.003$  & $6.182$   &    $25584\pm141$   &  $25.0\pm4.4$  \\
\sixiv &  $5.880\pm0.001$  & $6.182$   &    $21990\pm47$   &  $21.4\pm3.2$  \\
\enddata
\end{deluxetable}

\clearpage

\begin{deluxetable}{lccc}
\tabletypesize{\scriptsize}
\tablecaption{Best fit photoionization model parameters.}
\tablewidth{0pt}
\tablehead{
\colhead{} &  \colhead{MEG} & \multicolumn{2}{c}{HEG} \\
\cline{3-4}\\
\colhead{Parameter}  &\colhead{HV-LIP} & \colhead{HV-HIP-I} & 
\colhead{HV-HIP-II}
}
\startdata
$Log~U$  &   $-0.1\pm0.2$ & $2.5\pm0.2$ & $2.5\pm0.2$  \\
$Log~N_{H}~($cm$^{-2})$ & $20.2\pm0.2$ &  $23.5\pm0.3$ & $23.5\pm0.2$ \\
$V_{Turb}~($km~s$^{-1})$ & $303\pm41$ & $688\pm123$  &  $890\pm112$\\
$V_{Out}~($km~s$^{-1})$ & $52800\pm300$ &  $26010\pm210$ & $22140\pm120$ \\
$\Delta\chi^{2}\tablenotemark{a}$  & 11 &  24 & 10 \\
\enddata
\tablenotemark{a}{Improvement in $\chi^{2}$ to fit after adding model 
component}
\end{deluxetable}

\clearpage

\begin{figure}
\epsscale{.80}
\plotone{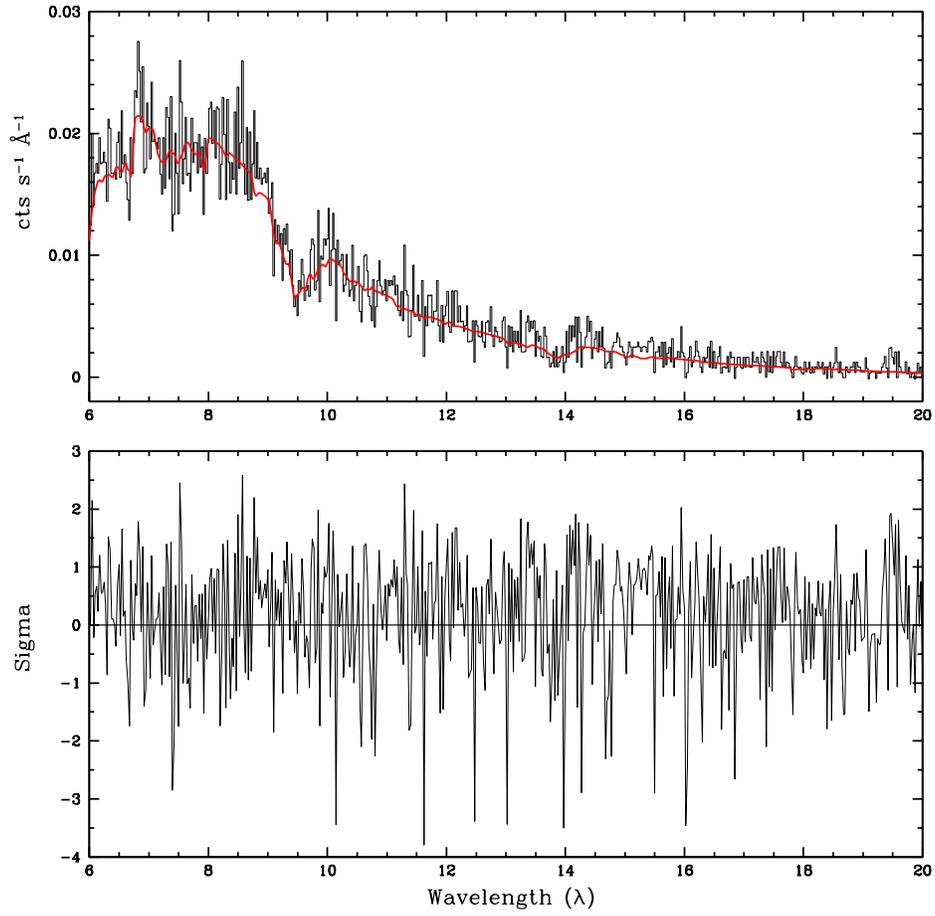}
\caption{The Mrk~590 MEG spectrum in the observer frame. The red line shows 
the best fit continuum model. In the bottom panel, plotted are the residuals 
showing absorber features.}
\end{figure}

\clearpage

\begin{figure}
\epsscale{.80}
\plotone{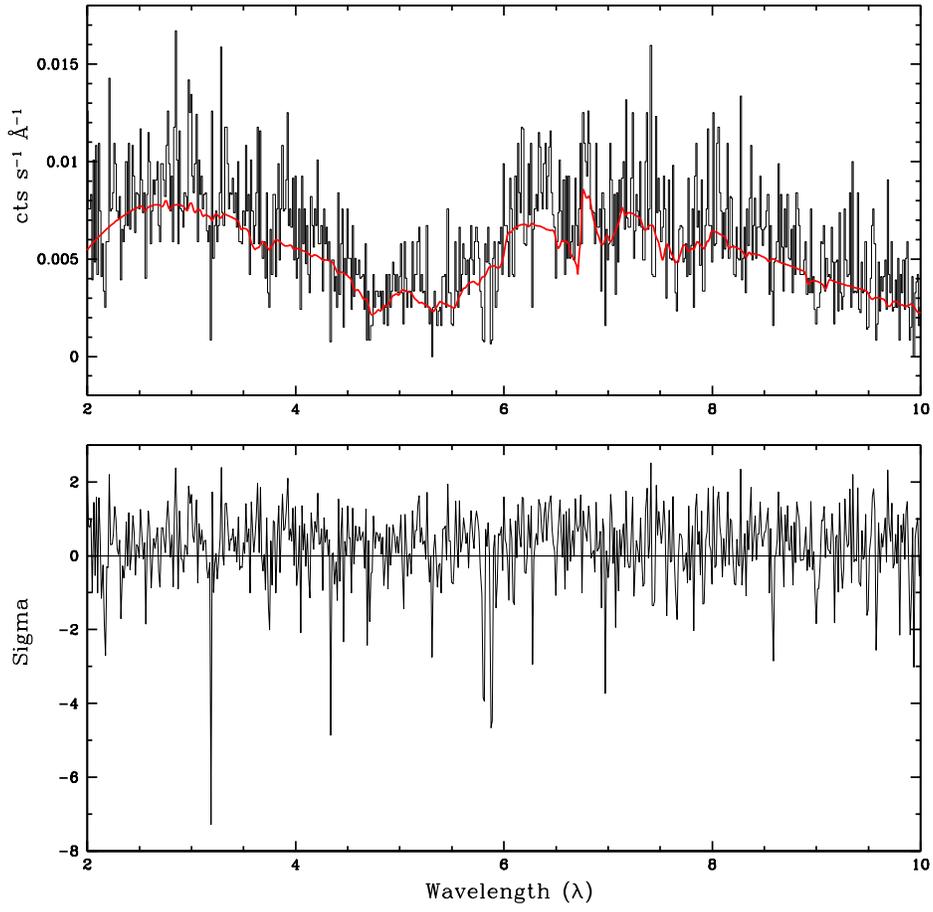}
\caption{Same as fig. 1 for HEG spectrum.}
\end{figure}

\clearpage

\begin{figure}
\epsscale{.80}
\plotone{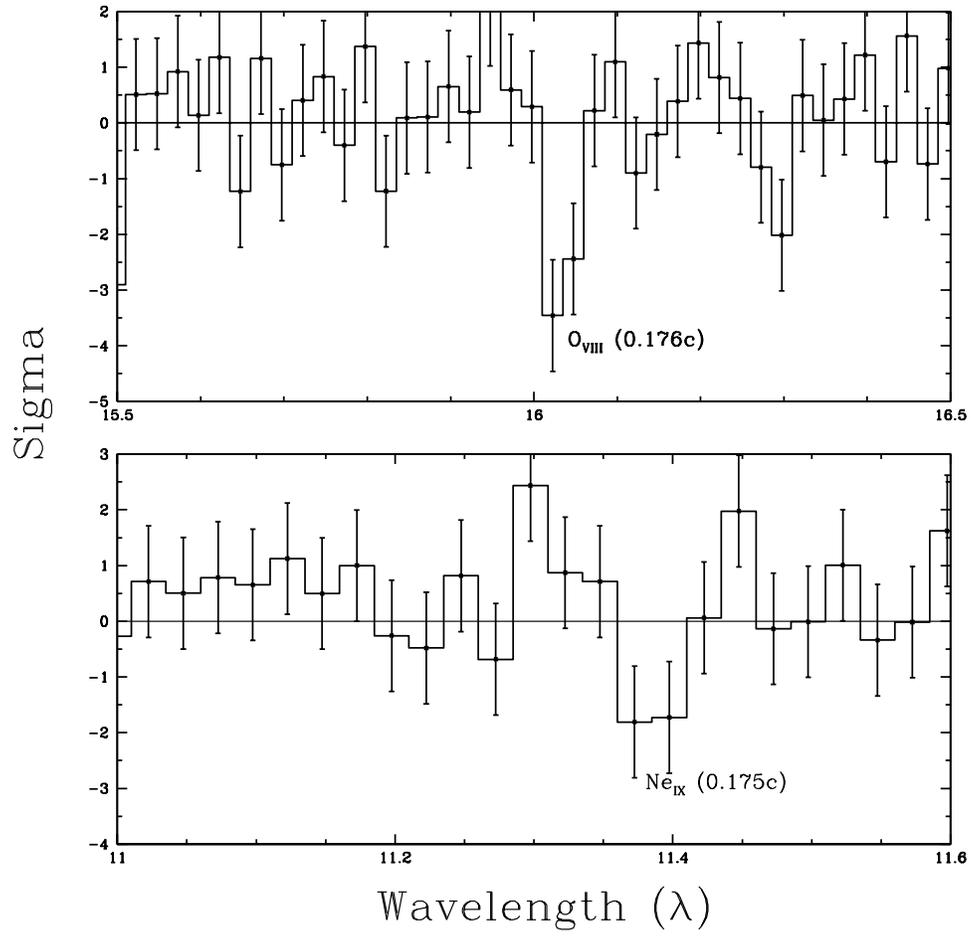}
\caption{Showing the high significance residuals of data:continuum fit to the 
MEG spectrum of Mrk~590, in the observer frame. }
\end{figure}

\clearpage

\begin{figure}
\epsscale{.80}
\plotone{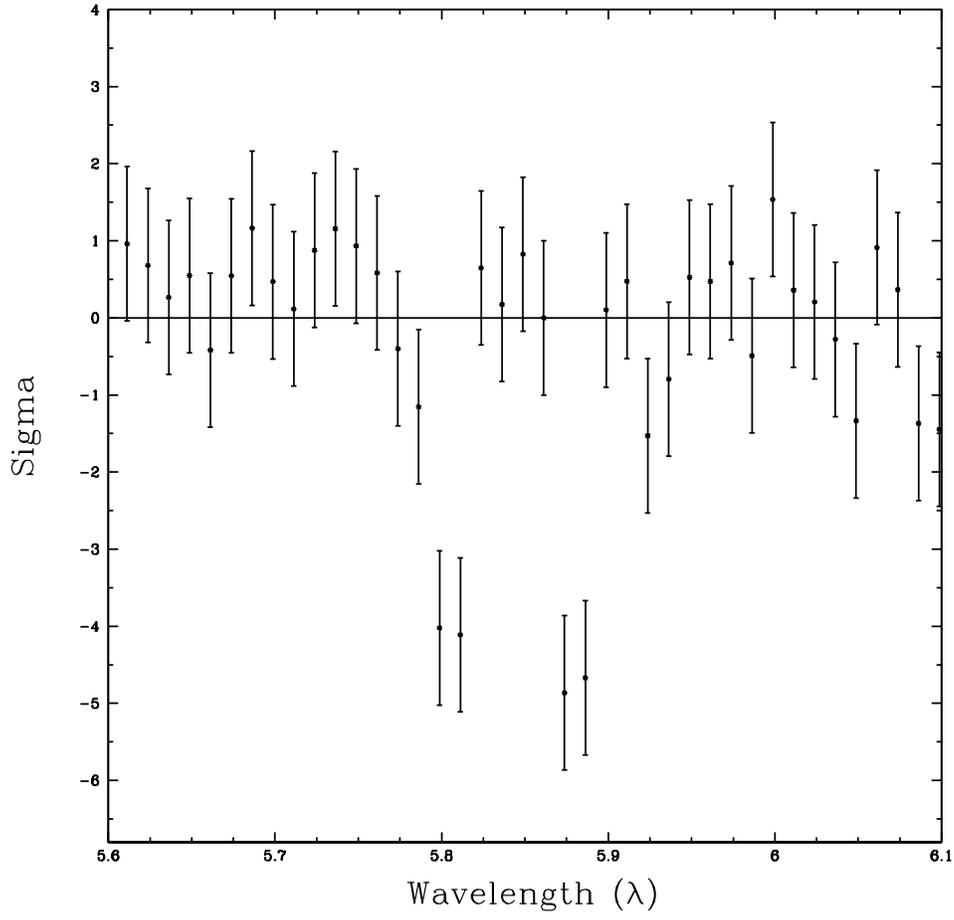}
\caption{ Same as fig. 3 for HEG data.}
\end{figure}

\clearpage

\begin{figure}
\epsscale{.80}
\plotone{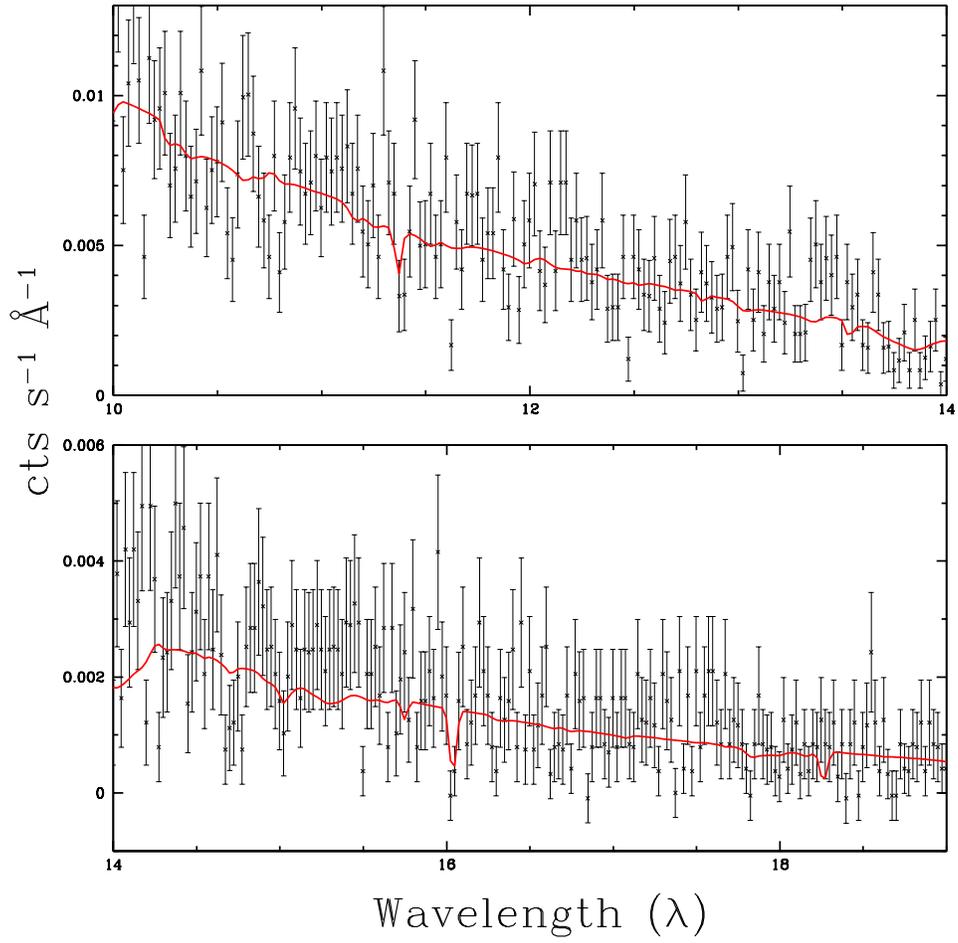}
\caption{The red solid line shows the best fit absorber model at 
outflow velocity of 0.176c.}
\end{figure}

\clearpage

\begin{figure}
\epsscale{.80}
\plotone{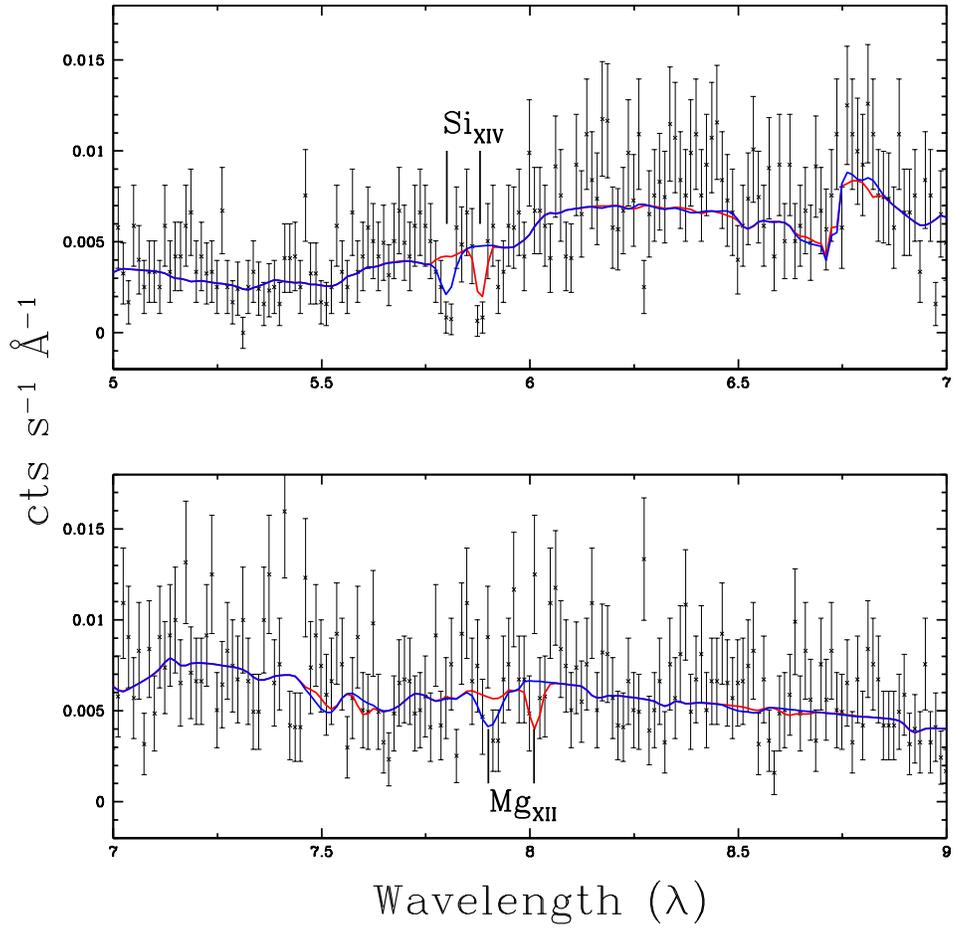}
\caption{The red and blue lines show the best fit absorber models at outflow 
velocities of $0.0867c$ and $0.0738c$, respectively.}
\end{figure}

\clearpage

\begin{figure}
\epsscale{.80}
\plotone{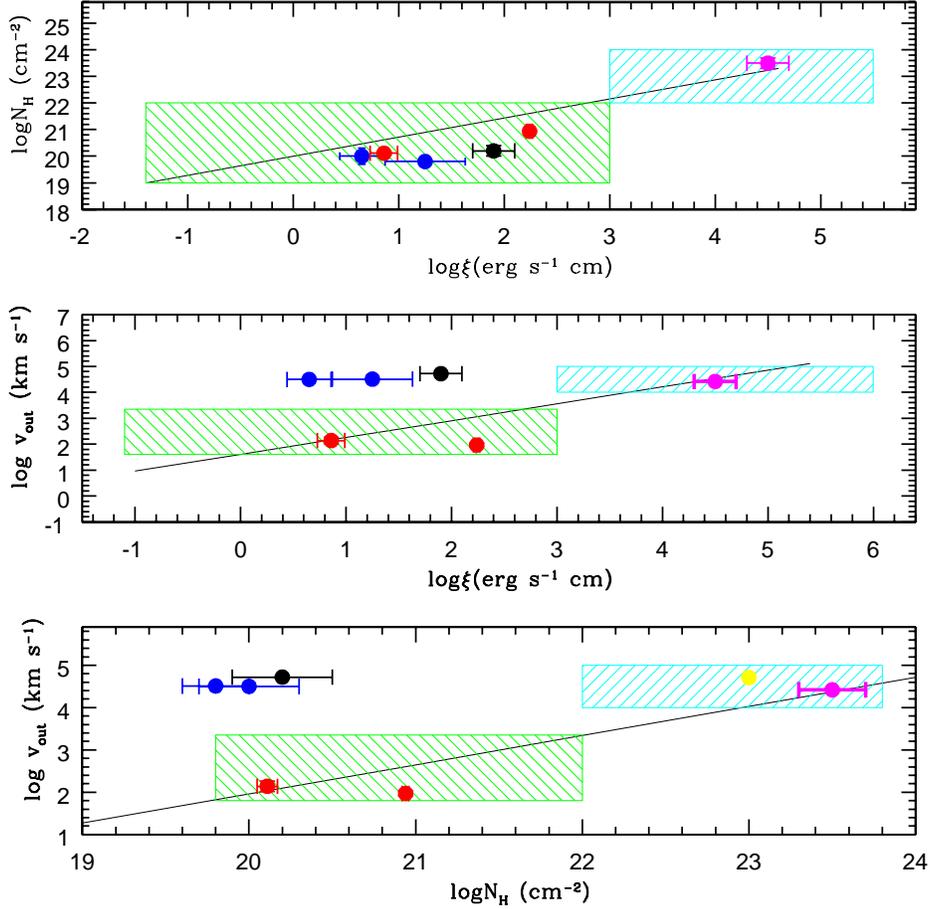}
\caption{\small The $\log \xi$ vs. $\log N_{H}$ (top panel), $\log \xi$
vs.  $\log v_{out}~$ (middle panel) and $\log N_{H}$ vs. $\log v_{out}~$
(bottom panel) for the low-velocity WAs (green striped region) and UFOs
(blue striped region) using data from Tombesi et al. (2013). The solid
lines represent the correlation fits to low-velocity WAs and UFOs from
Tombesi et al. The data-points represent the Mrk~590 and Ark~564  outflow 
parameters, (1) Mrk~590: LIP High-velocity outflows (black) and  
HIP high-velocity outflows (magenta)  
(2) Ark~564: low-velocity WAs (red; Gupta et al. 2013a), 
UFO (yellow; Papadakis et al. 2007) and high-velocity outflows 
(blue; Gupta et al. 2013b). As we see, the high-velocity low-ionization 
absorbers in Ark~564 and Mrk~590 occupy an unexplored region of the 
parameter space. }
\end{figure}

\end{document}